\documentclass[twocolumn,showpacs,preprintnumbers,amsmath,amssymb,superscriptaddress, prb, floatfix]{revtex4}
\usepackage{graphicx}% Include figure files
\usepackage{dcolumn}% Align table columns on decimal point
\usepackage{bm}% bold math
\usepackage{eucal}
\begin{document}

\newcommand{\bmu}{\mbox{\boldmath $\mu$ } }
\newcommand{\bJ}{\mbox{\boldmath $J$ } }
\newcommand{\bs}{\mbox{\boldmath $s$ } }
\newcommand{\bh}{\mbox{\boldmath $h$ } }
\newcommand{\bH}{\mbox{\boldmath $H$ } }
\newcommand{\bT}{\mbox{\boldmath $T$ } }\newcommand{\rZ}{ \rm Z  }
\newcommand{\rB}{ \rm B  }
\newcommand{\rCF}{ \rm CF  }
\newcommand{\rM}{ \rm M  }
\newcommand{\rQ}{ \rm Q  }

\preprint{APS/123-QED}

\title{Neutron Scattering Study on the Field-Induced ${O_{xy}}$-type Antiferroquadrupolar Ordering of Heavy-Fermion Superconductor PrOs${_{4}}$Sb${_{12}}$}
% Force line breaks with \\

\author{K. Kaneko}
\affiliation{Advanced Science Research Center, Japan Atomic Energy Agency, Tokai, Naka, Ibaraki 319-1195, Japan}
\email{kaneko.koji@jaea.go.jp}

\author{N. Metoki}
\affiliation{Advanced Science Research Center, Japan Atomic Energy Agency, Tokai, Naka, Ibaraki 319-1195, Japan}
\affiliation{Department of Physics, Tohoku University, Sendai 980-8578, Japan}

\author{R. Shiina}
\affiliation{Department of Physics, Tokyo Metropolitan University, Hachioji, Tokyo 192-0397, Japan}

\author{T. D. Matsuda}
\affiliation{Advanced Science Research Center, Japan Atomic Energy Agency, Tokai, Naka, Ibaraki 319-1195, Japan}

\author{M. Kohgi}
\affiliation{Department of Physics, Tokyo Metropolitan University, Hachioji, Tokyo 192-0397, Japan}

\author{K. Kuwahara}
\affiliation{Department of Physics, Tokyo Metropolitan University, Hachioji, Tokyo 192-0397, Japan}

\author{N. Bernhoeft}
\affiliation{DRFMC-CEA, 38054 Grenoble, France}
\altaffiliation{Present address: 18 Maynestone Road, SK23 6AQ, United Kingdom}

\date{\today}% It is always \today, today,

\begin{abstract}
Neutron scattering experiments have revealed a field-induced antiferroquadrupolar order parameter in the Pr-based heavy-fermion superconductor PrOs$_4$Sb$_{12}$.
We observed the field-induced antiferromagnetic dipole moment with the propagation vector {\textit{\textbf{q}}} = (1\,0\,0) for the applied field direction both {\textit{\textbf{H}}}${\parallel}$[1\,1\,0] and [0\,0\,1]. 
For {\textit{\textbf{H}}}${\parallel}$[1\,1\,0] at 8\,T, it should be noted that the induced antiferromagnetic moment of 0.16(10) ${\mu}_{\rm B}$/Pr orients parallel to the field within our experimental accuracy.
This observation is strong evidence that the $O_{xy}$ electric quadrupole to be the primary order parameter for {\textit{\textbf{H}}}${\parallel}$[1\,1\,0].
A mean-field calculation, based on ${\Gamma}_1$ singlet and ${\Gamma}_4^{(2)}$ excited triplet with the $O_{xy}$ quadrupolar interaction, reproduces the induced moment direction and its field response.
These facts indicate the predominance of $O_{xy}$-type antiferroquadrupolar interaction in PrOs$_4$Sb$_{12}$.
\end{abstract}

\maketitle

\section{\label{sec:level1}Introduction}
PrOs$_4$Sb$_{12}$ is the first Pr-based heavy-fermion superconductor and attracts considerable interest because of unusual superconducting properties.\cite{PrOsSbMBM_1,PrOsSbEDB_1,PrOsSbHS_1,PrOsSbTen_1,PrOsSbHK_1,PrOsSbKI_1,PrOsSbYA_2,PrOsSbTG_1,PrOsSb_GS_01}
One of the most interesting topics is what is the role of the multipole moments of Pr $f$ electrons for the heavy-fermion nature and unconventional superconductivity in PrOs$_4$Sb$_{12}$.\cite{PrOsSbTK_2, PrOsSbTT_2}
Long range magnetic dipole order does not coexist with superconductivity at zero field.\cite{PrOsSbMBM_1,PrOsSbEDB_1}
Instead, the existence of a magnetic field-induced ordered phase has been reported above the upper critical field $H_{\rm c2}$ for the superconductivity.\cite{PrOsSbYA_1,PrOsSbRV_1,PrOsSbTT_1,PrOsSbCRR_1}
In our previous study antiferromagnetic reflections with {\textit{\textbf{q}}}=(1\,0\,0) were observed in the field-induced ordered phase for {\textit{\textbf{H}}}${\parallel}$[0\,0\,1].\cite{PrOsSbMK_1}
The antiferromagnetic moment of 0.02\,${\mu}_{\rm B}$/Pr at 8\,T along the [0\,1\,0] direction perpendicular to the applied field was significantly smaller than the field-induced ferromagnetic moment parallel to {\textit{\textbf{H}}}. 
This small dipole moment and the $H$-$T$ phase diagram can be well explained by the mean-field calculation based on the crystal field level scheme of the ${\Gamma}_1$ singlet ground state with ${\Gamma}_{4}^{(2)}$ triplet excited state and assuming an antiferroquadrupolar (AFQ) interaction.\cite{PrOsSbMK_1, PrOsSbRS_1,PrOsSbRS_2}
The field-induced ordered phase appears as a result of the level crossing of the zero field ground and excited state which comes from the field split triplet state. 
The level scheme and its field response have been experimentally clarified by means of neutron inelastic scattering under magnetic fields.\cite{PrOsSbKKuwa_1}
The overall aspect of the field-induced ordered phase indicates that the antiferromagnetic dipole is not a primary order parameter.

Generally speaking, there are many possibilities for the primary multipole order parameters.
The ordered phase is stable only under applied magnetic field, where the time reversal symmetry has been broken, 
thus, the antiferromagnetic dipole order appears simultaneously with the coupled order parameters 
such as quadrupole, octupole, and hexadecapole in the order parameter space with given $T_h$ symmetry.\cite{PrOsSbRS_2}
Since neutrons exhibit no cross section for an electric quadrupole moment, the observation of the field-induced antiferromagnetic peak by neutron is not a direct proof for the primary AFQ order parameter. 
In this sense, our previous study did not rule out possibilities of other high-rank multipoles.\cite{PrOsSbMK_1}
Actually, it is shown that the qualitative features for {\textit{\textbf{H}}}${\parallel}$[0\,0\,1], the induced antiferromagnetic moment perpendicular to the applied field and the phase diagram can be reproduced in terms of octupole as well as quadrupole\cite{PrOsSbRS_2, PrOsSbRS_4}.
Recent intensive studies on skutterudites report an important role of higher rank multipole, octupole\cite{SmRuPMY_2} and hexadecapole\cite{PrRuPTT_1}, in some compounds. 
Therefore, it is indispensable to clarify strictly the dominant interaction in PrOs$_4$Sb$_{12}$.
For this purpose, further information is required: field direction and pressure dependencies and inelastic neutron scattering.

Careful magnetization measurements on PrOs$_4$Sb$_{12}$ clarified that the field-induced ordered phase also exists for [1\,1\,0] and [1\,1\,1] with remarkable anisotropy in the $H$-$T$ phase diagram.\cite{PrOsSbTT_1}
The existence of the field-induced ordered phase for these different field directions also supports the singlet-triplet crystal field level scheme. 
It has been pointed out that the anisotropy in the phase diagram is intrinsic to $T_h$ symmetry and the predominance of quadrupolar interaction.\cite{PrOsSbRS_1,PrOsSbRS_2}
The field-induced ordered states for both {\textit{\textbf{H}}}${\parallel}$[1\,1\,0] and [1\,1\,1] are also considered to be the AFQ ordered phase and not to be the dipolar one.
The order parameter for {\textit{\textbf{H}}}${\parallel}$[1\,1\,0] was theoretically predicted to be $O_{xy}$\cite{PrOsSbRS_2}, but has not been confirmed so far.
If this is the case, one could discriminate the primary order parameter of non-magnetic quadrupole from coupled magnetic octupole since the induced antiferromagnetic moment direction is expected to be different between them.
Therefore the study for {\textit{\textbf{H}}}${\parallel}$[1\,1\,0] is worthwhile in order to clarify a dominant interaction in PrOs$_4$Sb$_{12}$.
Besides, it motivates us that the magnetization measurements revealed the existence of some additional anomalies within the field-induced ordered phase for {\textit{\textbf{H}}}${\parallel}$[1\,1\,0], indicating possible transitions between multipolar ordered states. \cite{PrOsSbTT_1}

The purpose of the present neutron scattering study is to clarify the order parameter of the field-induced ordered phase for {\textit{\textbf{H}}}${\parallel}$[1\,1\,0] and to investigate the origin of anomalies in this ordered state. 
We found that the induced antiferromagnetic moment is parallel to the field direction, which is the strong evidence for the $O_{xy}$-type quadrupole as the primary order parameter.

\begin{figure}[t]
\includegraphics[width=4.0cm]{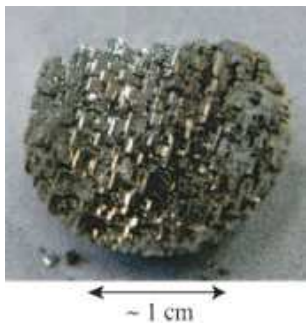}
\caption{{\label{f1}} 
A photograph of the single crystalline sample of PrOs$_4$Sb$_{12}$ used for the present neutron scattering study.}
\end{figure}%

\section{\label{sec:level2}Experiment}

\begin{figure}[t]
\includegraphics[width=8.0cm]{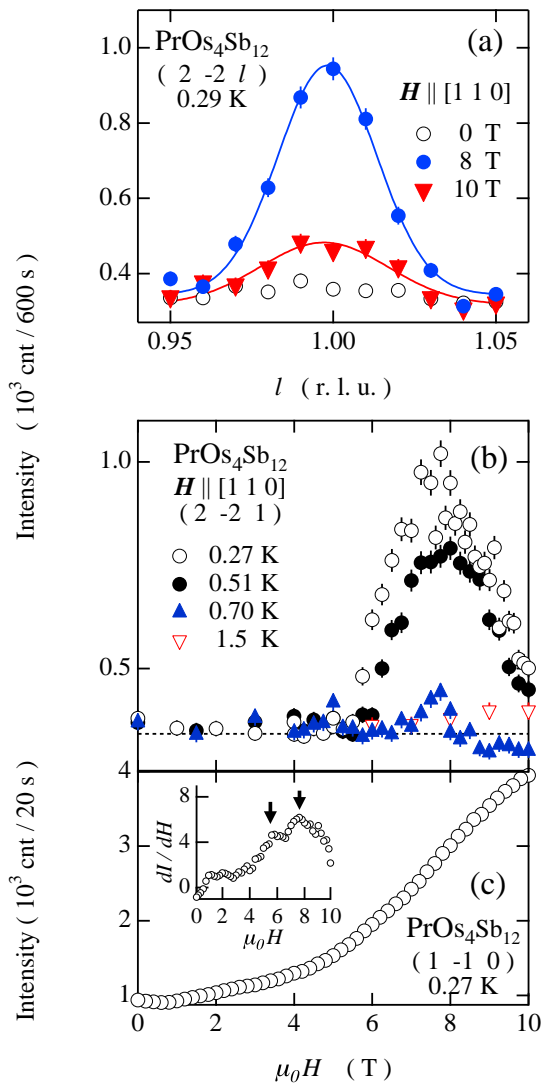}
\caption{{\label{f2}}
(a)	The field dependence of the 2\,\={2}\,1 superlattice reflection profile of PrOs$_4$Sb$_{12}$ measured at 0.29 K  for {\textit{\textbf{H}}}${\parallel}$[1\,1\,0]. 
The solid lines denote Gaussian fitting.
(b)	The field dependence of the 2\,\={2}\,1 reflection intensity measured at various temperatures. 
(c)	The 1\,\={1}\,0 reflection intensity plotted as a function of applied magnetic field taken at 0.27\,K. 
The inset gives its differential curve, d$I$/d$H$.}
\end{figure}%
Neutron scattering experiments have been carried out on the cold neutron triple-axis spectrometer LTAS installed in the guide hall of the research reactor JRR-3 in Japan Atomic Energy Agency, JAEA. 
The instrumental setup was just the same as our previous experiment.\cite{PrOsSbMK_1}
The use of a cold neutron with triple-axis mode was crucial for taking high quality data to reduce the contamination of elastic spectra by the low-energy excitations.
We used a liquid He-free 10\,T superconducting magnet and $^3$He-$^4$He dilution refrigerator both developed by JAEA.~\cite{10Tmag_SK_1}
The vertical field was applied along [1\,1\,0], perpendicular to the ($h$\,${\bar{h}}$\,$l$) scattering plane. 
We observed superlattice peaks at $h+k+l=odd$, while ferromagnetic scattering is superposed on nuclear Bragg peak at $h+k+l=even$, for example (1\,\={1}\,0).

A large single crystalline sample with a mass of 6\,g has been grown by the antimony-self-flux method. 
The details of the sample preparation have been published elsewhere.\cite{PrOsSbHS_1,PrOsSb_KK_2}
Figure~\ref{f1} shows the picture of the sample, which is composed of small crystallites typically 1 mm$^3$ distributed within 1 degree of the crystallographic axes. 
The four-circle x-ray crystal structure analysis on the small piece of this sample confirms the filled skutterudite structure with the space group $I{m}{\bar{3}}$ ($T_{h}^{5}$, \#204). 
The sample was characterized via magnetic susceptibility measurements; we observed clear Curie-Weiss behavior at high temperature and the maximum at 3.6\,K, which is identical to the high-quality sample used in the de Haas van Alphen study.\cite{PrOsSbHS_1}
The resistivity was also very similar to the previous results and our sample showed superconducting transition at the onset of 1.88\,K. 
These results guarantee the high quality of the present sample. 
Fortunately there was no domain structure concerning the two-fold symmetry in the basal plane, thus for example, ($h\,k\,l$) reflection is generally different from ($k\,h\,l$). 
We do not know the mechanism for the single domain growth process, nevertheless the single domain sample can be reproducibly obtained with our sample growth condition, which is rather mysterious.

\begin{figure}[t]
\includegraphics[width=8.0cm]{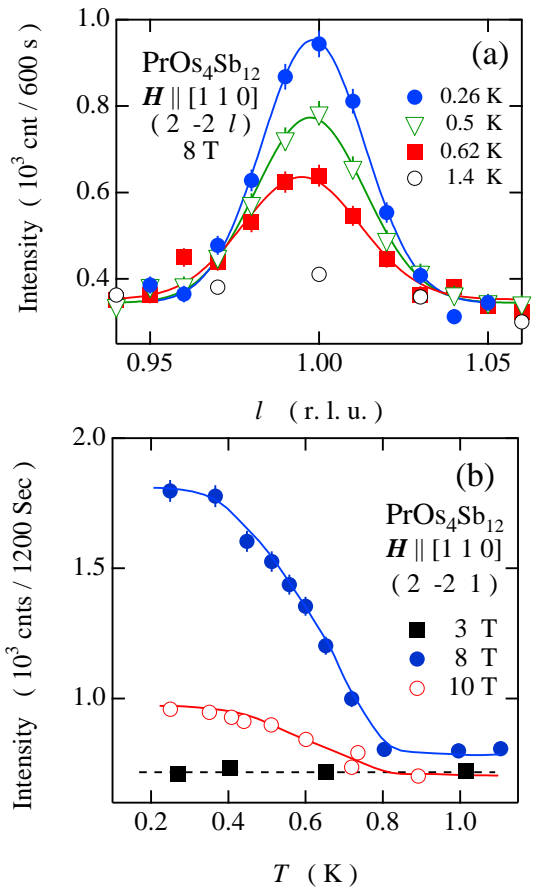}
\caption{{\label{f3}}
(a)	The temperature dependence of the 2\,\={2}\,1 superlattice reflection of PrOs$_4$Sb$_{12}$ measured under the magnetic field of 8\,T applied along the [1\,1\,0] direction. 
The solid lines denote the result of fitting with Gaussian and linear background for each temperature.
(b)	The temperature dependence of the 2\,\={2}\,1 reflection intensity measured at various fields. 
The solid lines are guides for the eyes.}
\end{figure}%
\begin{figure}[t]
\includegraphics[width=8.0cm]{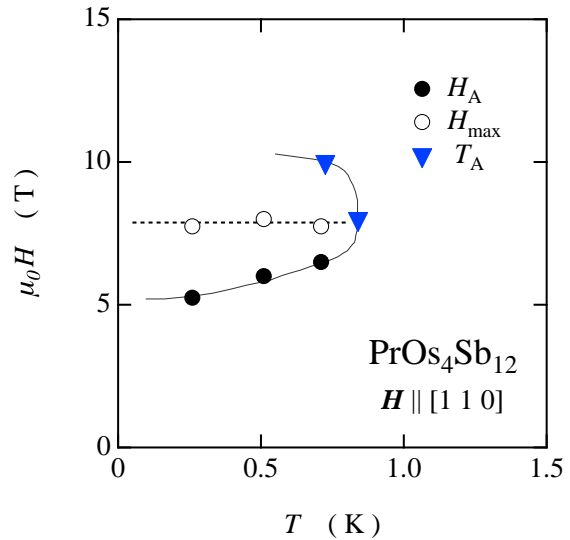}
\caption{{\label{f4}}
The $H$-$T$ phase diagram of PrOs$_4$Sb$_{12}$ for {\textit{\textbf{H}}}${\parallel}$[1\,1\,0]. 
The closed and open circles are fields corresponding to the onset and the maximum of the field-induced antiferromagnetic peak, respectively.
The closed triangles are the phase boundary determined from the temperature dependence. }
\end{figure}%

\section{\label{sec:level3}Results}
Figure~\ref{f2}(a) shows the representative data for the scattering profile of the 2\,\={2}\,1 superlattice peak measured at 0.29 K with magnetic fields along the [1\,1\,0] direction. 
An application of the magnetic field of 8\,T induces a clear resolution-limited superlattice reflection at (2\,\={2}\,1) where no trace of peak was observed at zero field. 
The observed field-induced superlattice peak positions are the same as those for {\textit{\textbf{H}}}${\parallel}$[0\,0\,1], namely the propagation vector is {\textit{\textbf{q}}}=(1\,0\,0). 
A marked decrease to ${\sim}$1/10 of the integrated intensity occurs between 8 and 10 T whereas  the magnetic Bragg peaks maintain their positions and widths.

The field dependence of the 2\,\=2\,1 peak intensity was measured at various temperatures as shown in Fig.~\ref{f2}(b).
Applying the field at 0.27\,K, the peak intensity appears at 5\,T (=$H_{\rm A}$), reaches a maximum around 8\,T (=$H_{\rm max}$), and decreases steeply above 8\,T.
$H_{\rm A}$ increases as the temperature rises: 6\,T  and 7\,T at 0.51\,K and 0.71\,K, respectively.
In contrast, the maximum around 8\,T is almost temperature independent.
No trace of superlattice peak was found above 1\,K.

Figure~\ref{f2}(c) displays a field variation of the peak intensity at 0.27\,K taken at {\textit{\textbf{Q}}}=(1\,${\bar 1}$\,0).
An increase of 1\,${\bar 1}$\,0 reflection intensity originating from the uniform magnetic moment was clearly observed.
The inset gives the differential curve, d$I$/d$H$, which corresponds to the differential curve of the square of the ferromagnetic moment $M^2$.
With applied fields, the intensity exhibits a gradual increase and changes its slope around 5.5\,T and 8\,T, which can be clearly seen in the inset.
These inflection fields are consistent with the onset  ($H_{\rm A}$) and maximum fields ($H_{\rm max}$) in the superlattice reflection intensity as well as the result of magnetization.\cite{PrOsSbTT_1}.
The small hump around 1.5\,T might arise from the breaking of the heavy-fermion superconductivity since it is very close to the upper critical field $H_{\rm c2}$.

\begin{figure}[t]
\includegraphics[width=8.0cm]{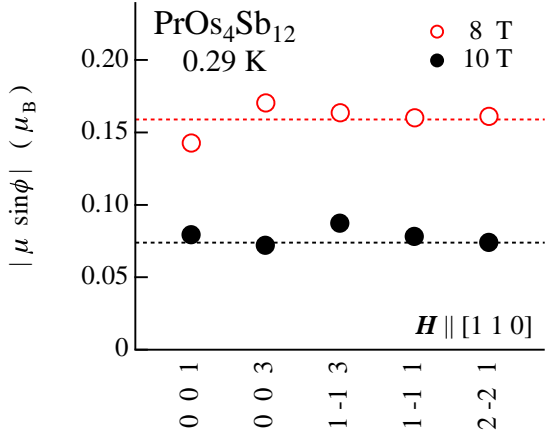}
\caption{{\label{f5}} 
The product of the magnitude of the magnetic moment and the angle factor derived from the observed magnetic Bragg reflection intensity.
The open and closed circles indicate the data for ${\mu}_0H=$8\,T and 10\,T, respectively.
The dotted lines are the fitting results for the reflection in the ($h {\bar h} l$) scattering plane by assuming the antiferromagnetic moment parallel to the applied field, namely, $\sin{\phi}=1$.}
\end{figure}%
The temperature dependence of the 2\,\={2}\,1 superlattice peak profile under the applied field of 8\,T is shown in Fig.~\ref{f3}(a). 
The 2\,\={2}\,1 superlattice reflection becomes weak without significant peak shift and/or broadening as temperature increases, and no trace of peak was observed at 1.4\,K.
Figure~\ref{f3}(b) shows peak intensity at (2\,\=2\,1) as a function of temperature taken under the field of 3, 8 and 10\,T.
With increasing temperature from 0.27\,K at 8\,T, the peak intensity showed the monotonous decrease and disappeared around  $T_{\rm A}$=0.8\,K.
The feature for 10\,T is quite similar to that for 8\,T except for the absolute intensity.
No peak was observed in the result for 3\,T.
No additional anomaly at 8\,T and 10\,T is consistent with the fact that $H_{\rm max}$ is almost temperature independent.			
The present results for {\textbf{\textit{H}}}${\parallel}$[1\,1\,0] of PrOs$_4$Sb$_{12}$ are summarized in the $H$-$T$ magnetic phase diagram as shown in Fig.~\ref{f4}.
The closed and open circles are the field corresponding to the onset and the maximum of the field-induced antiferromagnetic peak, respectively.
The closed triangles are the phase boundary determined from the temperature dependence of the antiferromagnetic intensity. 
The overall aspect of the phase boundary, determined from the present diffraction experiments, is consistent with the result of magnetization measurements.\cite{PrOsSbTT_1}
The field-induced superlattice reflection with {\textit{\textbf{q}}}=(1\,0\,0) exists only in the region of the field-induced ordered phase, in other words, the ordered state with {\textit{\textbf{q}}}=(1\,0\,0) is the basis of the field-induced ordered phase for {\textbf{\textit{H}}}${\parallel}$[1\,1\,0] which is the same as {\textbf{\textit{H}}}${\parallel}$[0\,0\,1].

\begin{figure}[t]
\includegraphics[width=8.0cm]{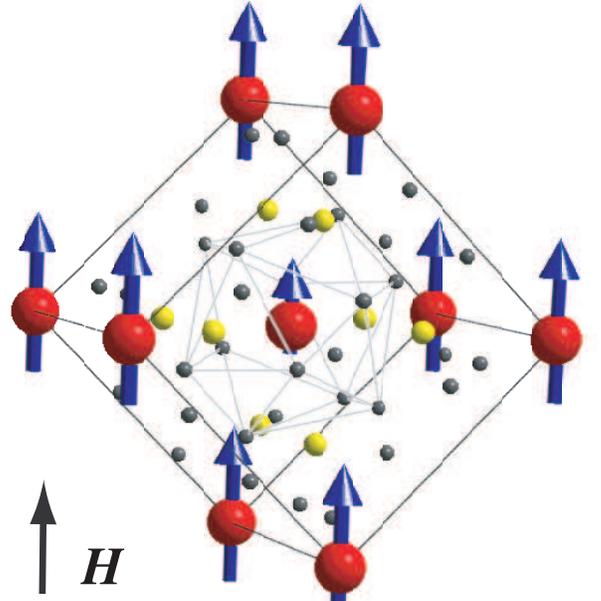}
\caption{{\label{f6}}
The obtained magnetic structure for the field-induced ordered phase of PrOs$_4$Sb$_{12}$ with application of the magnetic field along {\textit{\textbf{H}}}${\parallel}$[1\,1\,0].}
\end{figure}%
In order to determine the magnetic structure in the field-induced ordered phase, the integrated intensity of the superlattice peaks on the ($h\,\bar{h}\,l$) scattering plane were measured. 
The magnetic Bragg reflection intensity for unpolarized neutron diffraction can be written,
\begin{equation}
I_{\rm mag}(Q)=K L(\theta) [0.269\times10^{-12}f(Q)({\mu}\sin{\phi})\mid{F_{\rm mag}(Q)}\mid]^2
\label{eq:Nint}
\end{equation}
where $K$ is the scale factor, $L(\theta)$ is the Lorentz factor,  ${\mu}$ is the antiferromagnetic moment, $f(Q)$ is the magnetic form factor and $F_{\rm mag}(Q)$ is the magnetic structure factor which is unity for the present case.
${\phi}$ is the angle between the ordered magnetic moment and the scattering vector {\boldmath$Q$}.
The square of the product of angle factor and antiferromagnetic moment ${\mu}$ can be given as follows;
\begin{equation}
	\mid{\mu} \sin{\phi}\mid=\sqrt{\frac{I_{\rm mag}}{KL(\theta)[0.269\times10^{-12}f(Q)F_{\rm mag}(Q)]^2}}.
	\label{eq1}
\end{equation}%
Figure~\ref{f5} shows the square root of integrated intensity divided by scale factor, Lorentz factor, structure factor and magnetic form factor of the Pr$^{3+}$ free ion. 
Therefore the vertical axis in Fig.~\ref{f5} is the product of the antiferromagnetic moment ${\mu}$ and the angle factor $\sin{\phi}$. 
As clearly seen in Fig.~\ref{f5}, this quantity is almost isotropic within the present scattering plane. 
Although the data obtained at 10\,T is much weaker, it is also isotropic.
This isotropic nature of the left hand side of equation Eq.\,(\ref{eq1}) indicates that the antiferromagnetic moment is perpendicular to the ($h\,\bar{h}\,l$) scattering plane at both 8\,T and 10\,T within our experimental accuracy. 
Thus the induced dipole moment observed by neutron is parallel to the direction of the applied field along [1\,1\,0]. 
Using this angle factor, $\sin{\phi}$=1,  the induced antiferromagnetic moment for 8\,T and 10\,T is deduced to be ${\mu}_{\rm AF}=0.16(10)\,{\mu}_{\rm B}$ and $0.07\,{\mu}_{\rm B}$, respectively.
The antiferromagnetic moment exhibits a strong reduction in the magnitude from 8\,T to 10\,T.

There is a large induced ferromagnetic moment parallel to the field, which is roughly ten times larger than that of the field-induced antiferromagnetic moment.~\cite{PrOsSbTT_1}
The sum of the ferromagnetic and antiferromagnetic moment gives the magnetic structure in the field-induced ordered phase as shown in Fig.~\ref{f6}. 
The moment size is different for the each magnetic sublattice due to the antiferromagnetic component parallel to the field direction; the magnetic moment at the corner of the unit cell is larger than that at the center, or vice versa.

\section{\label{sec:level4}Mean-Field Analysis}
We present a mean-field calculation for the present results with {\textit{\textbf{H}}}${\parallel}$[1\,1\,0].
In PrOs$_4$Sb$_{12}$, the ${\Gamma}_1$ singlet ground state with the ${\Gamma}_4^{(2)}$ first excited triplet state at 8\,K is well separated from the other excited state.\cite{PrOsSbMK_1, PrOs4Sb12_KK_04, PrOs4Sb12_NAF_02}
This means that  the low-lying singlet-triplet levels are responsible for low temperature physical properties.
Hence, we move on our discussions to this low-lying singlet-triplet subspace.
Active multipoles for the singlet-triplet system with {\textit{\textbf{J}}}=4 in the $O_h$ and $T_h$ symmetries are classified in ref.\,\onlinecite{PrOsSbRS_1} and \onlinecite{PrOsSbRS_2}.
According to the symmetry lowering from $O_h$ to $T_h$, the triplet states in $T_h$ are represented by linear combinations of those in $O_h$,
whereas the ${\Gamma}_1$ singlet is not influenced.
The mixing is parameterized by $y$, the coefficient in the crystal field Hamiltonian reflecting the effect of $T_h$, where $y=0$ corresponds to $O_h$.\cite{PrOsSbKT_1}

The possible multipole order parameters under magnetic fields which cause admixture are classified by symmetry analysis.
There are two possible symmetries for {\textit{\textbf{H}}}${\parallel}$[1\,1\,0], ${\Gamma}_1$ and ${\Gamma}_2$.
For ${\Gamma}_1$, the magnetic dipole $J_x+J_y$ is mixed with $J_x-J_y$ depending on the magnitude of $y$.
The induced antiferromagnetic moment in the field-induced ordered phase was revealed to be $J_x+J_y$ in the present experiment.
This result indicates the order parameter for {\textit{\textbf{H}}}${\parallel}$[1\,1\,0] to be the ${\Gamma}_1$ symmetry which is consistent with the theoretical prediction.\cite{PrOsSbRS_2}
Furthermore, the induced moment parallel to applied fields within the experimental accuracy suggests the small $y$ parameter which is also consistent with the previous results on neutron diffraction for {\textit{\textbf{H}}}${\parallel}$[0\,0\,1] and a small anisotropy in the $H$-$T$ phase diagram.
\begin{figure}[t]
\includegraphics[width=8.0cm]{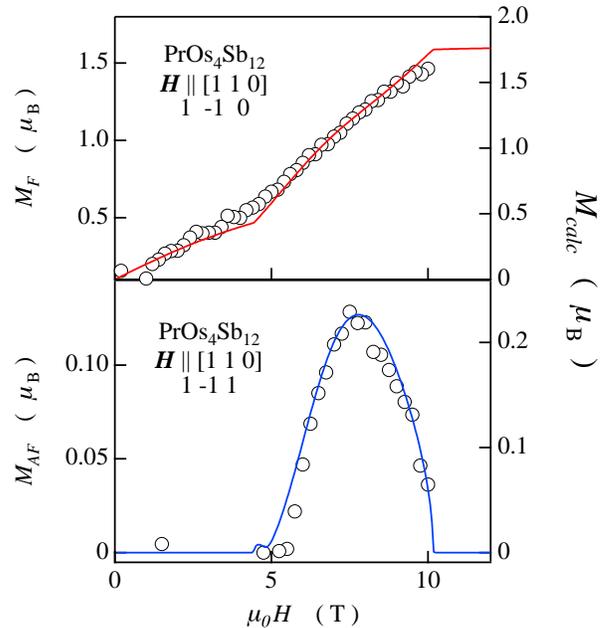}
\caption{{\label{f7}}
The field dependence of the field-induced ferromagnetic and antiferromagnetic moment. The open circles are obtained from the neutron diffraction, whose value corresponds to the left axes. The results of mean-field calculation are represented by lines corresponding to the right axes. }
\end{figure}%

Hereafter, we study to what extent the simple quadrupolar interaction model can reproduce the present observed results. 
Our analysis is based on the Hamiltonian consisting of crystal field potential, Zeeman energy and AFQ interaction, %
\begin{equation}
	H = H_{\rCF} + H_{\rZ} + H_{\rQ}. 
\end{equation} %
It is shown that the mechanism of field-induced AFQ order in this model can be clarified with a pseudo-spin representation analogous to the dimer-spin systems.\cite{PrOsSbRS_1, PrOsSbRS_2}
With the use of two pseudo-spins, $s_1$ and $s_2$ with $S=1/2$, which are defined in the singlet-triplet subspace,  the local part of the Hamiltonian is described as
\begin{eqnarray}
	H_{\rCF} & + & H_{\rZ} =\Delta \sum_i \left( \frac{3}{4} + \bs_{1i} \cdot \bs_{2i}\right)
	\nonumber \\
	& - & \bh \cdot \sum_i
	\big( \bs_{1i}+\bs_{2i} + \delta ( \bs_{1i}-\bs_{2i} ) \big),
\end{eqnarray}%
where $\Delta$ is the singlet-triplet energy splitting and $\bh$ is a scaled magnetic field. 
The appearance of an effective staggered field ($\delta$) to the two spins is characteristic of the $T_h$ system. 
It has been shown that the interaction of $O_{xy}$-type quadrupoles is mapped to the biquadratic pseudo-spin Hamiltonian, %
\begin{align}
	H_{\rQ} & = D_{\rQ} \sum_{(ij)} 4 \Big[   ( \bs_{1i} \times \bs_{2i}  ) \cdot ( \bs_{1j} \times \bs_{2j}  )+ \epsilon_1 \bmu_i \cdot \bmu_j \nonumber \\ 
	& + \epsilon_2  \Big(  ( \bs_{1i} \times \bs_{2i}  ) \cdot \bmu_j + \bmu_i \cdot ( \bs_{1j} \times \bs_{2j}  ) \Big) \Big], 
\end{align}%
where%
\begin{equation}
	\bmu =( s_1^y s_2^z+s_1^z s_2^y, \, s_1^z s_2^x+s_1^x s_2^z, \, s_1^x s_2^y+s_1^y s_2^x ). 
\end{equation}%
See ref. \onlinecite{PrOsSbRS_1} and \onlinecite{PrOsSbRS_2} for details on the dependence of $\epsilon_1$ and $\epsilon_2$ on the $T_h$ parameter $y$. 
An advantage of this mapping is to clarify that the predominant and isotropic contribution to the field-induced order is given by the part of a vector product $\bs_1 \times \bs_2 $, whereas the remaining part related with $\bmu$ produces an anisotropic correction due to the $T_h$ symmetry. 

In the following, we discuss the mean-field solutions of this model in the field direction {\textit{\textbf{H}}}${\parallel}[1\,1\,0]$, in which the stable phase is shown to be of $O_{xy}$. 
The calculation has been carried out by using the same parameter set as in ref. \onlinecite{PrOsSbRS_2}. 
Namely, we assume $D_{\rQ}z = 0.3 \Delta$,  $\Delta=8$K, $x=0.45$ and $y=0.12$, where $z$ is the number of the nearest neighbor bonds, and $(x,y)$ are the crystal field parameters.
The results of magnetic field dependence of the ferromagnetic ($M_{\rm F}$) and antiferromagnetic moment ($M_{\rm AF}$) are shown in Fig.~\ref{f7}. 
They are compared with the present experimental results that are deduced by normalizing $M_{\rm F}$ and $M_{\rm AF}$ at 8\,T to be 1.2$\mu_{\rB}$\cite{PrOsSbTT_1} and 0.16$\mu_{\rB}$, respectively. 
One finds that the observed overall field response of both ferromagnetic and antiferromagnetic moments is well explained by this model analysis. 
These agreement clearly indicates that the staggered moment should be induced by the primary AFQ order which remains inaccessible to the neutron probe. 
It is also shown that the calculated value of antiferromagnetic moment is larger than that for $H{\parallel}[0\,0\,1]$, and this tendency is qualitatively consistent with the observed results. 
Note, however, that there exist quantitative differences between theory and experiment, probably due to fluctuation effects or contribution of other multipole interactions. 
  
\begin{figure}[t]
\includegraphics[width=8.0cm]{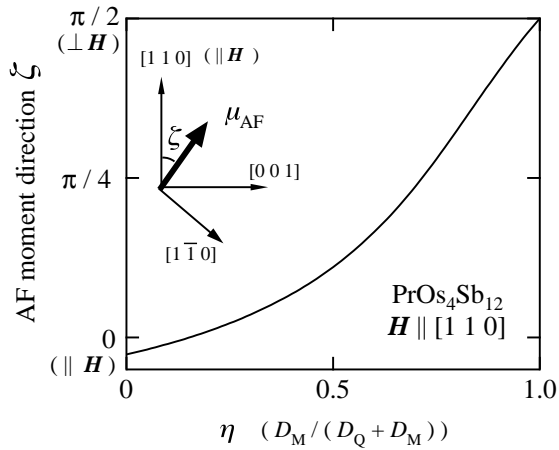}
\caption{{\label{f8}} 
Calculated angle ${\zeta}$ between the field-induced antiferromagnetic moment and the field direction [1\,1\,0] as a function of relative magnetic interaction strength ${\eta}$ for $h/{\Delta}$=1.2, corresponding to ${\mu}_0H{\simeq}$7\,T.}
\end{figure}%
In order to clarify the unique properties of the AFQ model more definitely, 
let us study the effects of possible magnetic interaction. 
It is shown that the magnetic interaction of dipoles and octupoles is expressed as an isotropic bilinear spin Hamiltonian.\cite{PrOsSbRS_2}
Although the generic form of the interaction is somewhat complex with a few parameters, we introduce here the simplest form to capture the essential physics, 
\begin{align}
H_{\rM} = D_{\rM} \sum_{(ij)} ( \bs_{1i} \cdot \bs_{1j} + \bs_{2i} \cdot \bs_{2j} ).  
\end{align} %
This symmetric model in the two spins corresponds to a mixed interaction of dipoles and octupoles. 
Note, however, that the relevant part is the octupolar interaction, because they have much larger matrix elements between singlet and triplet states. 
It should be stressed that the total Hamiltonian in the limit, $D_{\rQ} = 0$ and $D_{\rM} \neq 0 $, reduces to the conventional dimer-spin model\cite{Haldane_MT_1},
in which a similar field-induced order takes place by the magnetic interaction. 
Thus, it is quite instructive to see the difference between quadrupolar and magnetic interactions by interpolating the two limits.

In the framework of the mean-field theory, we have calculated the antiferromagnetic moment direction for the total Hamiltonian $H+H_{\rM}$ in the field direction {\textit{\textbf{H}}}${\parallel}[1\,1\,0]$. 
Taking the ratio of the interactions as $D_{\rQ}=D(1-\eta)$ and $D_{\rM}=D\eta$, 
the angle from the applied field direction, ${\zeta}$, is plotted in Fig.~\ref{f8} as a function of relative magnetic strength $\eta=D_{\rM}/(D_{\rQ}+D_{\rM})$ with a fixed $Dz(=0.3\Delta)$. 
In the case that the magnetic interaction is small, the moment direction is almost parallel to the field. 
Whereas the magnetic interaction becomes dominant, the moment orients perpendicular to the field, as expected in the dimer-spin systems. 
Such a remarkable change of the antiferromagnetic structure that depends on the interaction is characteristic in the field direction {\textit{\textbf{H}}}${\parallel}[1\,1\,0]$. 
In the case of {\textit{\textbf{H}}}${\parallel}[0\,0\,1]$, both interactions result in the antiferromagnetic moment perpendicular to the field.\cite{PrOsSbRS_5} 
Thus, the present experiment for {\textit{\textbf{H}}}${\parallel}[1\,1\,0]$ is crucial in distinguishing the type of multipolar interactions. 
In this sense, the observed antiferromagnetic moment parallel to the field is regarded as strong evidence of the predominant AFQ interaction. 

\section{\label{sec:level5}Discussion}
The field-induced antiferromagnetic moment parallel to the field for {\textit{\textbf{H}}}${\parallel}$[1\,1\,0] is different from the perpendicular antiferromagnetic moment for {\textit{\textbf{H}}}${\parallel}$[0\,0\,1], whereas the propagation vector {\boldmath$q$}=(1\,0\,0) is the same. 
In the latter case the antiferromagnetic moment is parallel to the [0\,1\,0] direction, and not to the [1\,0\,0] direction; 
a clear peak was observed for {\textit{\textbf{q}}}=(1\,0\,0) but no trace of the antiferromagntic peak was observed for {\textit{\textbf{q}}}=(0\,1\,0). 
This characteristic moment direction reflecting the lack of four-fold symmetry gives the important information on the possible order parameter.
Variation in the orientation of the field-induced antiferromagnetic moment according to the applied field direction is characteristic to the quadrupolar ordered phase.
Both results indicate the predominance of the $O_{xy}$-type AFQ interaction in PrOs$_4$Sb$_{12}$. 

A recent inelastic neutron scattering study clarified that the crystal field excitations from the ${\Gamma}_1$ ground state to the ${\Gamma}_{4}^{(2)}$ triplet first excited state have a dispersion; the observed excitation softens at {\textit{\textbf{q}}} = (1\,0\,0) which is the same as the ordering vector of the field-induced AFQ ordered state.\cite{PrOs4Sb12_KK_04}
In addition, the intensity at {\textit{\textbf{q}}}=(1\,0\,0) is weaker than that at zone center.
Though the magnetic and non-magnetic interaction gives the same energy dispersion, the {\textit{\textbf{q}}} dependences of the intensity are opposite;\cite{PrOs4Sb12_RS_03} namely, magnetic interaction should lead to the stronger intensity at the zone boundary.
Thus, the observed excitation spectra indicate the dominance of antiferroquadrupolar interaction which is quite consistent with the present result for {\textit{\textbf{H}}}${\parallel}$[1\,1\,0].
Together with the anisotropy in the $H$-$T$ phase diagram, the predominance of AFQ interaction is clarified in PrOs$_4$Sb$_{12}$.

Furthermore, the excitation observed in the inelastic scattering study gives an interaction strength.
Note that the quadrupolar coupling constant obtained from the inelastic scattering is almost consistent with that used in the present study.
There is a slight difference in the definition; $D_{\rQ}$ in the present paper can be connected with $d_{\rQ}$ in ref.\onlinecite{PrOs4Sb12_KK_04} with $D_{\rQ}= {\beta}^2d_{\rQ}$ where ${\beta}$ is the off-diagonal matrix elements of quadrupolar moments between singlet and triplet.
Taking this difference into account, the coupling constant ${\beta}^2d_{\rQ}z{\simeq}$-2.7\,K derived from inelastic spectra is consistent with $D_{\rQ}z {\simeq}$2.4\,K used in the present study.
In addition, the quadrupolar coupling constant $g'_{{\Gamma}4}$ obtained from the ultrasound measurements of $C_{44}$ is also consistent with the present value.\cite{PrOsSbTG_1}
In other words, the coupling constant used in the present study corresponds with those obtained in the other measurements.

The maximum in the induced antiferromagnetic moment, $H_{\rm max}$, corresponds well to the $H_1$ anomaly in the magnetization\cite{PrOsSbTT_1}, as shown in Fig.~\ref{f4}.
However, no change in the antiferromagnetic structure was found on passing through $H_{\rm max}$.
As shown in Fig.~\ref{f7}, the induced $J_x+J_y$ appears to vanish slightly above 10\,T.
This critical field is well below the reported critical field for the field-induced ordered phase of ${\sim}$12\,T but rather close to the $H_2$ anomaly.\cite{PrOsSbTT_1}
In other words, there is another field-induced ordered state above $H_2$.
A possible explanation would be the switching of the order parameter from ${\Gamma}_1$ to ${\Gamma}_2$ representation at $H_2$. 
It is theoretically predicted that the energy difference of these two ordered states is very small.\cite{PrOsSbRS_2}
The coupled magnetic dipole order parameter of ${\Gamma}_2$ representation is $J_z$ for {\textit{\textbf{H}}}${\parallel}$[1\,1\,0]. 
Thus a change of moment direction from [1\,1\,0]${\parallel}H$ to [0\,0\,1]${\perp}H$ should be accompanied by the quadrupole-quadrupole transition. 
This reorientation of the antiferomagnetic moment has not been observed in our present neutron scattering experiment up to 10\,T.
In order to clarify the origin of the anomalies observed in magnetization, higher field neutron scattering experiments should be carried out.

In conclusion, we observed the field-induced antiferromagnetic order of heavy-fermion superconductor PrOs$_4$Sb$_{12}$ for {\textit{\textbf{H}}}${\parallel}$[1\,1\,0]. 
The antiferromagnetic moment was parallel to the field direction, which is direct evidence for the primary quadrupolar order parameter of the field-induced ordered phase. 
Together with our previous study for {\textit{\textbf{H}}}${\parallel}$[0\,0\,1], we can conclude that the $O_{xy}$-type quadrupolar interaction is intrinsic to the many body interaction between $f$ electron based quasiparticles in heavy-fermion superconductor PrOs$_4$Sb$_{12}$.  

\begin{acknowledgments}
We are very grateful to J. -M. Mignot and A. Gukasov for communication of unpublished data obtained at LLB Saclay.
We also wish to thank H. Sugawara for helpful advice on sample preparation.
This work was supported by Grants-in-Aid for Scientific Research, Young Scientist (B) (No. 16740212) and in Priority Area "Skutterudite" (No. 18027012, 18027013 and 18027015) of the Ministry of Education, Culture , Sports, Science  and Technology, Japan.
\end{acknowledgments}


\begin{thebibliography}{99}
\bibitem{PrOsSbRS_5}
In the case of the $O_h$ symmetry ($y=0$), the dipoles and the octupoles in the singlet-triplet system are simply expressed with the pseudo spins as $\bJ \propto \bs_1+\bs_2$ 
and $\bT^\beta \propto \bs_1 - \bs_2 $.\cite{PrOsSbRS_2}
When the $T_h$ term is introduced ($y \neq 0$), they are weakly mixed with each other. 
Concerning the mean-field solution of the interaction (7), it is known that the antiferromangetic order parameter is transverse to the field direction\cite{Haldane_MT_1}.
If the $y$ component of the spins, or equivalently that of the dipoles and the octupoles, is chosen as the order parameter in $\bH{\parallel}[0\,0\,1]$, one can expect 
$\langle s_1^y \rangle \sim - \langle s_2^y \rangle $ due to the positive $\Delta$ term in eq. (4). 
This leads to a tiny antiferrmagnetism because $J_y \sim s_1^y + s_2^y$ for a small $T_h$ term, which is compatible to the experimental results for $\bH{\parallel}[0\,0\,1]$. 
Since the model is isotropic for the field directions, this transverse antiferromagnetism does not account for the present experimental results in $\bH{\parallel}[1\,1\,0]$.
\end{thebibliography}
\end{document}